\documentclass{cimento}

\usepackage{graphicx}  
\usepackage{cite}

\title{Disentangling electroweak effects in $Z$-boson production}

\author{S.~Carrazza\from{ins:x}}
\instlist{\inst{ins:x} Dipartimento di Fisica, Universit\`a di Milano \& INFN, Sezione di Milano}

\PACSes{\PACSit{12.15.Lk}{Electroweak radiative corrections}}

\begin{document}

\maketitle

\begin{abstract}
  Parton distributions with QED corrections open new scenarios for
  high precision physics. We recall the need for accurate and improved
  predictions which keeps into account higher order QCD corrections
  together with electroweak effects. We study predictions obtained
  with the improved Born approximation and the $G_{\mu}$ scheme by
  using two public codes: {\tt DYNNLO} and {\tt HORACE}. We focus our
  attention on the Drell-Yan $Z$-boson invariant mass distribution at
  low- and high-mass regions, recently measured by the ATLAS
  experiment and we estimate the impact of each component of the final
  prediction. We show that electroweak corrections are larger than PDF
  uncertainties for modern PDF sets and therefore such corrections are
  necessary to improve the extraction of future PDF sets.
\end{abstract}

\section{Introduction}

Recently the NNPDF Collaboration published sets of parton distribution
functions with QED corrections, the so called {\tt NNPDF2.3QED} sets
\cite{Ball:2013hta,Bertone:2013vaa,Carrazza:2013wua,Carrazza:2013bra,Carrazza:2013axa,Skands:2014pea}. These
sets contain the photon PDF with its uncertainty determined for the
first time from DIS and Drell-Yan LHC data.

In this work we estimate and compare to the PDF uncertainties the
contributions to the invariant mass of the Drell-Yan $Z$-boson
production due to electroweak corrections and the photon-induced
channel, by considering the low-mass region, which is below the $Z$
peak resonance and the high-mass tail.

In contrast to what was shown in Ref.~\cite{Boughezal:2013cwa} where
predictions were computed with {\tt FEWZ}, here we propose to combine
two distinct parton level public codes: {\tt
  DYNNLO}~\cite{Catani:2007vq} for the NLO QCD prediction and {\tt
  HORACE}~\cite{CarloniCalame:2007cd} which provides the exact
$\mathcal{O}(\alpha)$ electroweak radiative correction together with
the photon-induced channel for the $Z$ production.  The motivation for
this combination is the interest to measure the difference between
predictions with electroweak effects at NLO/NNLO QCD accuracy computed
in the improved Born approximation (IBA) instead of using electroweak
correction computed by {\tt FEWZ} in the $G_{\mu}$ scheme. The main
difference between these choices is that effective couplings in the
IBA reabsorb higher-order electroweak corrections and therefore it
provides predictions in better agreement with experimental data.

Computations are performed exclusively with the {\tt
  NNPDF23\_nlo\_as\_0119\_qed} set of PDFs instead of using the
respective LO and NNLO sets because here we will focus only on the NLO
QCD accuracy and that is why we use a NLO set.

In the next sections, we first show the differences at Born level
between the improved Born approximation (IBA), available in {\tt
  HORACE} by default, and the $G_{\mu}$ scheme in {\tt DYNNLO}, then,
we proceed with the construction of the full prediction.

\section{Comparing the improved Born approximation (IBA) with the $G_{\mu}$ scheme}

\begin{figure}
  \begin{centering}
    \includegraphics[scale=0.35]{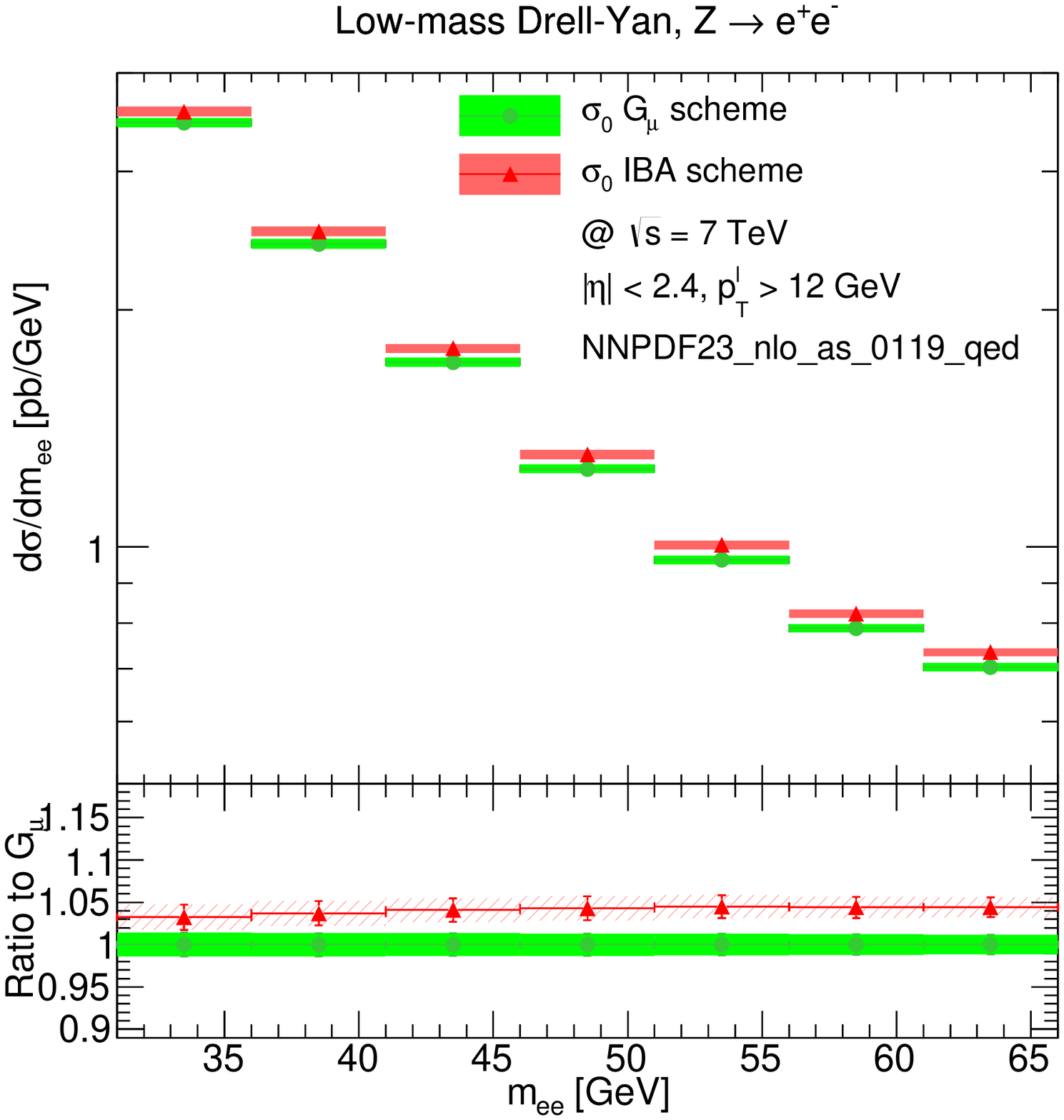}\includegraphics[scale=0.35]{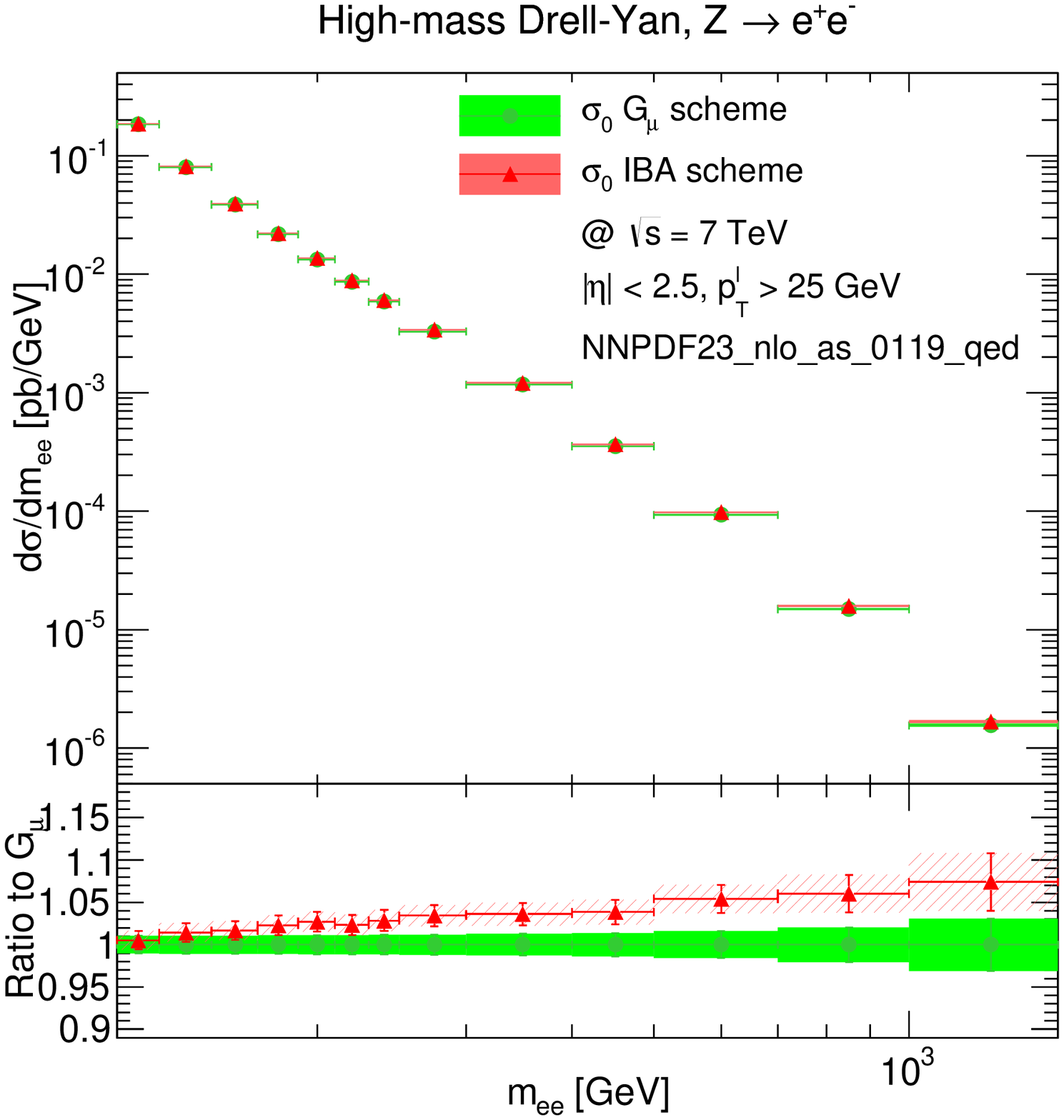}
    \par\end{centering}
    \caption{\label{fig:iba} Born level predictions and respective
      ratios for low- (left) and high-mass (right) Drell-Yan, $Z
      \rightarrow e^{+}e^{-}$ production, using the IBA and the
      $G_{\mu}$ scheme. At low-mass there is a constant gap of 3-4\%
      for all bins, while at high-mass, predictions increase
      progressively with the invariant mass, producing discrepancies
      of 7-8\% in the last bin.}
\end{figure}

In order to obtain realistic results, which are ready for comparisons
with real data, we have selected the kinematic range and cuts inspired
by recent measurements performed by the ATLAS experiment for low- and
high-mass Drell-Yan differential cross-section at $\sqrt{s}=7$ TeV
\cite{Aad:2013iua,Aad:2014qja}.

Figure~\ref{fig:iba} shows the predicted distribution at Born level
using the IBA ({\tt HORACE}) and the $G_{\mu}$ scheme ({\tt DYNNLO})
at low (left plot) and high (right plot) invariant mass regions, for
the Drell-Yan process: $Z \rightarrow e^{+}e^{-}$. Here, the goal is
to measure the numerical differences due to the choice of these
methodologies.

For all distributions, the Monte Carlo uncertainty is below the
percent level. We have computed predictions with the {\tt
  NNPDF23\_nlo\_as\_0119\_qed} set of PDFs because this is the set
that we use to build the complete prediction at NLO in QCD with
electroweak effects. The uncertainties shown in the figure have been
calculated as the 1-$\sigma$ interval obtained after averaging over
the 100 replicas provided by this set.

In the low-mass region, we have applied kinematic cuts to the lepton
pair imposing: $p_{T}^{l} > 12$ GeV and $|\eta^{l}| < 2.4$ as in
ATLAS~\cite{Aad:2014qja}. In this region we observe an almost flat gap
of 3-4\% between the IBA and $G_{\mu}$ predictions, however in the bin
$m_{ee}=51-56$ GeV the difference is slightly higher.

On the other hand, in the high-mass region we have applied the
following kinematic cuts: $p_{T}^{l} > 25$ GeV and $|\eta^{l}| < 2.5$
as in Ref.~\cite{Aad:2013iua}. We observe a progressive increase of
the central value prediction as a function of the invariant mass,
reaching a maximum of 7-8\% at the highest bin in $m_{ee}$. This
suggests that the running of $\alpha(Q^{2})$ in the IBA can play a
crucial role when determining with accuracy the predictions in such
region.

It is important to highlight that in both cases, PDF uncertainties are
smaller than the observed differences induced by the choice of the
scheme. These results are fully consistent with the IBA implementation
discussed in Ref.~\cite{CarloniCalame:2007cd}. In the sequel we are
interested in combining electroweak effects with higher order QCD
corrections in the IBA and then compare these results to pure QCD
$G_{\mu}$ predictions.

\section{Disentangling electroweak effects}

\begin{figure}
  \begin{centering}
    \includegraphics[scale=0.35]{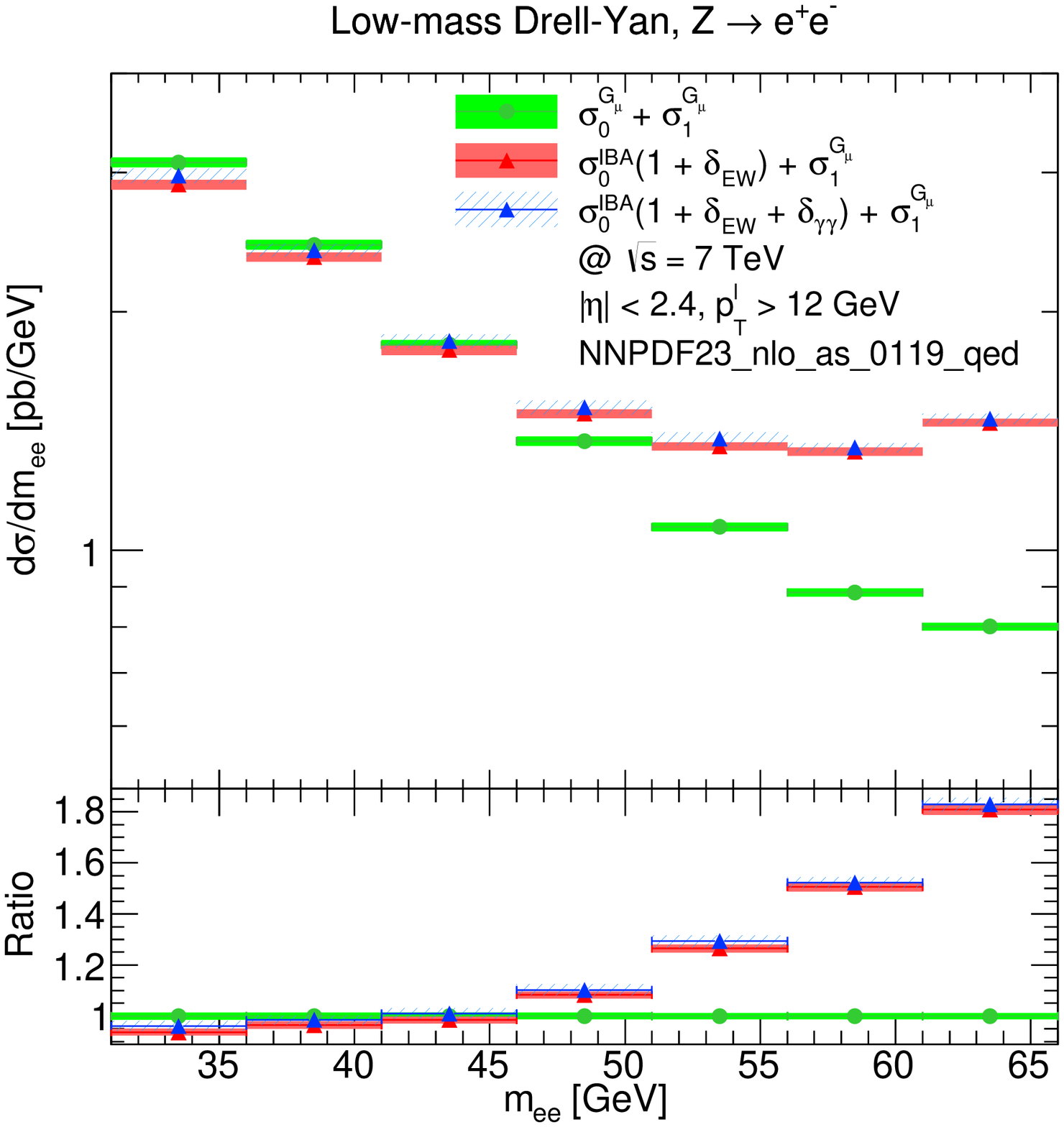}\includegraphics[scale=0.35]{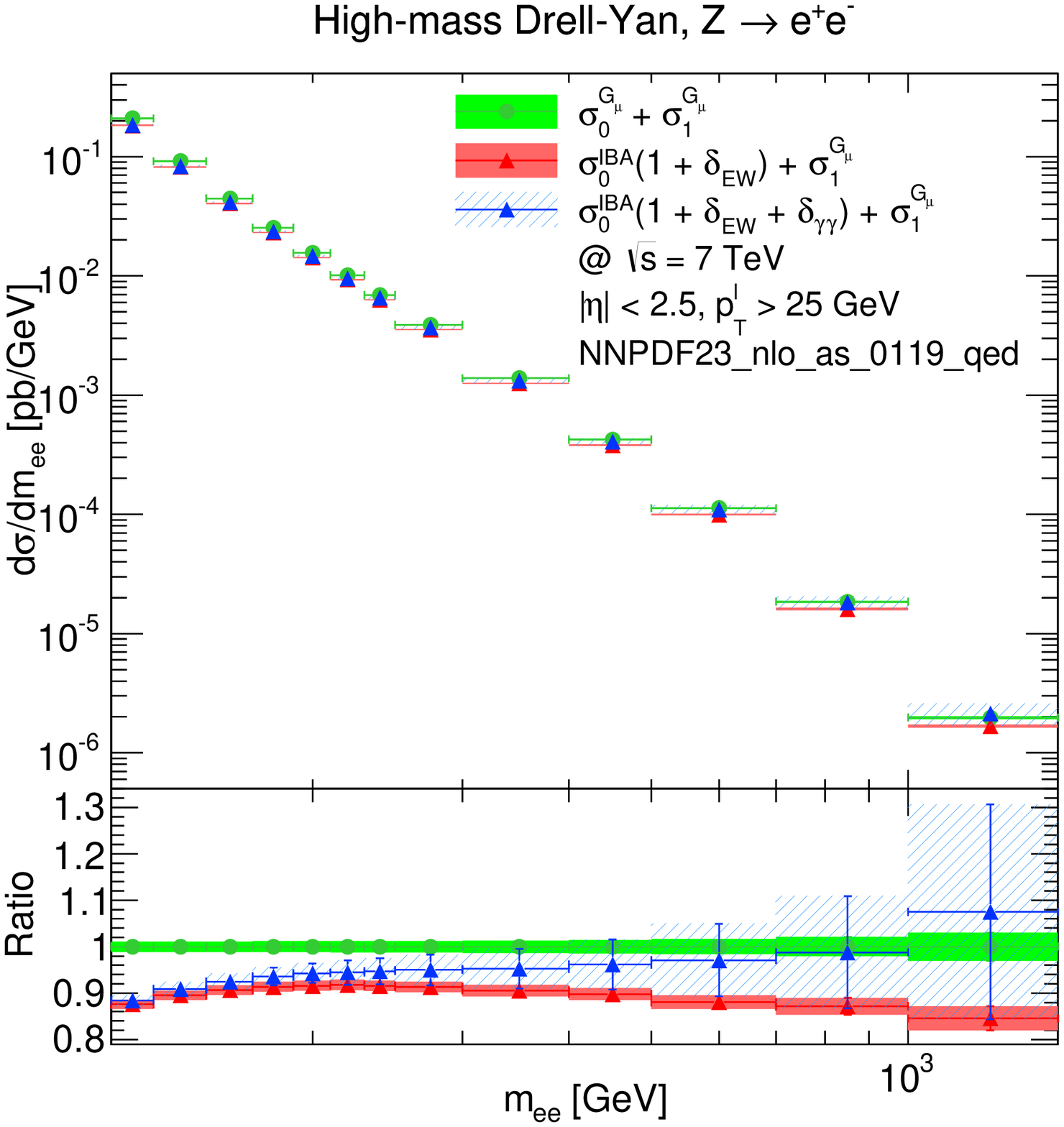}
    \par\end{centering}
    \caption{\label{fig:full} Comparison of predictions and respective
      ratios for low- (left) and high-mass (right) Drell-Yan, $Z
      \rightarrow e^{+}e^{-}$ production. We compare the NLO QCD
      prediction provided by {\tt DYNNLO} (green distribution) with:
      the combined prediction with $\delta_{\rm EW}$ (red
      distribution) and with the $\delta_{\rm EW} +
      \delta_{\gamma\gamma}$ (blue distribution).}
\end{figure}

At this point, we are interested in building a prediction based on IBA
which includes NLO QCD with $\mathcal{O}(\alpha)$ correction and the
photon-induced channel. We propose to extract the NLO correction from
{\tt DYNNLO} by removing its Born level, which contains the direct and
strong dependence on the $G_{\mu}$ scheme, and combine the result with
the {\tt HORACE} prediction. Schematically this can be achieved by
defining the quantities:
\begin{eqnarray}
  \sigma_{\rm {\tt DYNNLO}} = \sigma^{G_{\mu}}_{0} +
  \sigma^{G_{\mu}}_{1},\\ \sigma_{\rm {\tt HORACE}} = \sigma^{\rm IBA}_{0}
  (1+\delta_{\rm EW}+\delta_{\rm \gamma\gamma}),
\end{eqnarray} 
where $\sigma^{\rm IBA}_{0}$ and $\sigma^{G_{\mu}}_{0}$ are the Born
levels presented in Figure~\ref{fig:iba}, $\sigma_{1}^{G_{\mu}}$ the
NLO QCD, $\delta_{\rm EW}$ the $\mathcal{O}(\alpha)$ electroweak
correction and $\delta_{\gamma\gamma}$ the photon-induced
contribution.

The combination is then constructed in the following way:
\begin{eqnarray}
  \sigma_{\rm Total} & = &
  \sigma_{\rm {\tt DYNNLO}} + \sigma_{\rm {\tt HORACE}} - \sigma^{G_{\mu}}_{0}
  \\ & = & \sigma^{\rm IBA}_{0} + \sigma^{\rm IBA}_{0}
  \delta_{\rm EW}+\sigma^{\rm IBA}_{0} \delta_{\rm \gamma\gamma} +
  \sigma^{G_{\mu}}_{1}.\label{eq:1}
\end{eqnarray} 
where we remove the {\tt DYNNLO} Born level while we include the NLO
QCD correction in the final prediction.

We are aware that using this methodology we improve the combination
but we do not remove entirely the pure $G_{\mu}$ dependence at
higher orders, however this is the best combination we can propose
without applying technical modifications to both codes \footnote{In
  preparation by S.C., G. Ferrera, A. Vicini.}.

In Figure~\ref{fig:full} we compare $\sigma_{\rm {\tt DYNNLO}}$ with
$\sigma_{\rm Total}$, the combination presented in Eq.~\ref{eq:1},
with and without the $\delta_{\gamma\gamma}$ term. For all
distributions we compute the 1-$\sigma$ uncertainty except when
including the photon-induced channel where we have used the 68\%
c.l. as in Ref.~\cite{Ball:2013hta}.

In the low-mass region the inclusion of $\mathcal{O}(\alpha)$
electroweak corrections has a strong impact on the last four bins,
where differences can reach $\sim80\%$ in comparison to the pure NLO
QCD $G_{\mu}$ prediction, while the same correction for the high-mass
distribution shows a moderate impact which is below $\sim20\%$ for the
highest invariant mass bin. This behavior is expected and derives from
the shape of the $Z$-boson invariant mass: bins located in a region
lower than the $Z$ peak resonance undergoes large positive corrections
while at high invariant mass we observe a change of sign of such
corrections. It is important to highlight that modern data provided by
the LHC experiments are already corrected by final-state photon
radiation which carries a dominant fraction of the electroweak effects
shown in Figure~\ref{fig:full}.

The photon-induced contribution has a moderate impact in the low-mass
region while for high-mass it is dominant: this behavior is expected
and due to the presence of the $Z$ peak resonance where the
photon-induced channel is negligible.

Also from these plots of Figure~\ref{fig:full}, it is important to
emphasize again that modern PDF sets, as the {\tt NNPDF2.3QED}, have
uncertainties which are accurate enough to appreciate the differences
due to scheme choices and electroweak effects, including the new
photon PDF, which shows a stable behavior of uncertainties in all
invariant mass regions except at very high-mass bins where
uncertainties grow, reaching more than $\sim20\%$. This situation will
be improved in future by including more relevant and precise data to
constrain the photon PDF.

\section{Outlook}

We have shown that the choice of the coupling scheme clearly modifies
the predictions when looking at realistic experimental regions for the
$Z$-boson production. We have constructed a combined prediction which
includes the $\mathcal{O}(\alpha)$ electroweak corrections together
with the photon-induced channel in the IBA using public codes. We show
that uncertainties from modern PDF sets are smaller than electroweak
effects and therefore such corrections will provide noticeable
differences when extracting future PDF sets.

Finally, in a future work we expect to provide a similar study for
other processes and kinematic regions, by propagating consistently the
IBA to higher orders and producing a direct comparison to the {\tt
  FEWZ} results.


\begin{thebibliography}{0}

\bibitem{Ball:2013hta}
  R.~D.~Ball {\it et al.}  [NNPDF Collaboration],
  Nucl.\ Phys.\ B {\bf 877} (2013) 2,  290
  [arXiv:1308.0598 [hep-ph]].

\bibitem{Bertone:2013vaa}
  V.~Bertone, S.~Carrazza and J.~Rojo,
  arXiv:1310.1394 [hep-ph].

\bibitem{Carrazza:2013wua}
  S.~Carrazza [NNPDF Collaboration],
  arXiv:1305.4179 [hep-ph].

\bibitem{Carrazza:2013bra}
  S.~Carrazza [NNPDF Collaboration],
  PoS DIS {\bf 2013} (2013) 279
  [arXiv:1307.1131 [hep-ph]].

\bibitem{Carrazza:2013axa}
  S.~Carrazza, S.~Forte and J.~Rojo,
  arXiv:1311.5887 [hep-ph].

\bibitem{Skands:2014pea}
  P.~Skands, S.~Carrazza and J.~Rojo,
  arXiv:1404.5630 [hep-ph].

\bibitem{Boughezal:2013cwa}
  R.~Boughezal, Y.~Li and F.~Petriello,
  Phys.\ Rev.\ D {\bf 89} (2014) 034030
  [arXiv:1312.3972 [hep-ph]].


\bibitem{Catani:2007vq}
  S.~Catani and M.~Grazzini,
  Phys.\ Rev.\ Lett.\  {\bf 98} (2007) 222002
  [hep-ph/0703012].

\bibitem{CarloniCalame:2007cd}
  C.~M.~Carloni Calame, G.~Montagna, O.~Nicrosini and A.~Vicini,
  JHEP {\bf 0710} (2007) 109
  [arXiv:0710.1722 [hep-ph]].

\bibitem{Aad:2014qja}
  G.~Aad {\it et al.}  [ ATLAS Collaboration],
  arXiv:1404.1212 [hep-ex].

\bibitem{Aad:2013iua}
  G.~Aad {\it et al.}  [ATLAS Collaboration],
  Phys.\ Lett.\ B {\bf 725} (2013) 223
  [arXiv:1305.4192 [hep-ex]].

\end{thebibliography}
\end{document}